\documentclass[showpacs,preprintnumbers,amsmath,amssymb,axodraw,nofootinbib,floatfix, axodraw4j]{revtex4}

\usepackage{graphics}
\usepackage{axodraw}
\usepackage[dvips]{color}
\usepackage{amsmath}
\usepackage{graphicx}
\usepackage{epsf}
\usepackage{psfrag}
\usepackage{epsfig}
\usepackage[dvips]{color}
\newcommand{\pslash}{\hbox{$p\!\!\!{\slash}$}}

\newcommand{\kslash}{\hbox{$k\!\!\!{\slash}$}}

\newcommand{\be}{\begin{equation}}
\newcommand{\ee}{\end{equation}}
\newcommand{\bq}{\begin{eqnarray}}
\newcommand{\eq}{\end{eqnarray}}

\begin{document}

	\title{Gauge embedding procedure: classical and quantum equivalence between dual models}

	\date{\today}
	\author{B. Alves Marques$^{(a,b)}$} \email[] {bruno.alves@ifmg.edu.br}
	\author{B. Z. Felippe$^{(a,c)}$} \email[] {brunofelippe@unifei.edu.br}
	\author{A. P. Baeta Scarpelli$^{(a)}$} \email[] {scarpelli@cefetmg.br}
	\author{L. C. T. Brito$^{(d)}$} \email[]{lcbrito@ufla.br}
	\author{A. Yu. Petrov$^{(e)}$} \email[] {petrov@fisica.ufpb.br}
	
	\affiliation{(a)Centro Federal de Educa\c{c}\~ao Tecnol\'ogica - MG \\
		Avenida Amazonas, 7675 - 30510-000 - Nova Gameleira - Belo Horizonte
		-MG - Brazil}
	
	\affiliation{(b) Instituto Federal de Minas Gerais - Campus Sabará \\
		Rodovia MGC 262, Sobradinho, Minas Gerais, Brazil}
	
	\affiliation{(c) Universidade Federal de Itajubá - Campus Itabira \\
		Rua Irmã Ivone Drumond, 200 - Distrito Industrial II, Itabira, Minas Gerais, Brazil}
	
	\affiliation{(d) Universidade Federal de Lavras - Departamento de F\'{\i}sica \\
		Caixa Postal 3037, 37.200-000, Lavras, Minas Gerais, Brazil}
	
	\affiliation{(e) Departamento de F\'{\i}sica, Universidade Federal da Para\'{\i}ba\\
		Caixa Postal 5008, 58051-970, Jo\~ao Pessoa, Para\'{\i}ba, Brazil}

	\begin{abstract}
		\noindent
		In this paper the gauge embedding procedure of dualization is reassessed through a deeper analysis of the mutual equivalence of vector field models of more generic forms, explicitly, a general modified massive gauge-breaking extension of electrodynamics and its dual gauge-invariant model we derive in the paper. General relations between the vector field propagators and interaction terms of these models are obtained. Further, these models are shown to be equivalent at tree-level and one-loop physical calculations. Finally, we discuss extension of this equivalence to all loop orders.
		
	\end{abstract}
	
	\pacs{11.15.-q, 11.10.Kk, 11.10.Ef}

	\maketitle
	\section{Introduction}
	The concept of duality has been explored for several decades in different contexts. It is basically a mapping of the structures of models that, at first glance, seem to be distinct, but which, after a deeper analysis, turn out to be equivalent. We can say that a duality corresponds to equivalent descriptions of a model by using distinct Lagrangian formulations (for reviews, see \cite{review-1} and \cite{review-2}). One of the most interesting aspects of duality, with important implications in the perturbative context, concerns the exchange of the coupling regime from $g$ to $1/g$. In this case, the duality relation allows the mapping of a weak coupling in a strong coupling and vice versa, as in the case of the implementation of the relation between electric and magnetic couplings. It is important to note that duality can relate a theory without gauge invariance to a gauge invariant theory.
	
	In view of these relevant features, different methods have been developed to establish and investigate duality. We highlight especially the master action approach \cite{Malacarne}, whose starting point is an action depending on two fields, from which, two dual models are obtained by eliminating one field with the use of its field equations. The gauge embedding procedure \cite{ilha-1}, \cite{ilha-2}, also known as Noether Dualization Method (NDM), was developed to handle difficulties when dealing with the non-Abelian version of the self-dual model \cite{nonab1}, \cite{nonab2}, \cite{nonab3}. The well known equivalence in $2+1$ dimensions between the self-dual \cite{Townsend} and the topological massive \cite{topological} models, first shown by Deser and Jackiw \cite{Deser}, is easily obtained with the use of the gauge embedding approach. 
	
	The procedure of NDM is carried out by transforming the self-dual model into a gauge theory with the addition of on-shell vanishing terms. It is an iterative embedding of Noether counterterms, which is based on the idea of local lifting a global symmetry and which follows the concepts used in the papers by Freedman and van Nieuwenhuizen \cite{Nieuw2} and subsequent works by Ferrara, Freedman and van Nieuwenhuizen \cite{Ferrara1} and Ferrara and Scherk \cite{Ferrara2}, which played an important role for constructing the component-field supergravity actions.
	
	The gauge embedding method has been shown to be an efficient tool in the study of different field theory models. In particular, it allows to find new couplings for vector fields, as in the case of the self-dual theory minimally coupled to the spinor matter, which has been shown to generate the magnetic (nonminimal) and the Thirring-like current-current couplings \cite{ilha-2}. Further, with the cited methods, dualities have been established between the nonlinear generalizations of self-dual and Maxwell-Chern-Simons theories \cite{nlin}, in higher-rank tensor generalizations of these theories \cite{highrank} and in their higher-derivative extensions \cite{HD}. Besides, noncommutative extensions of the duality have been discussed in \cite{NC}. There were also several works elaborated extending duality to Lorentz-violating (LV) models, like for the LV extension of the 3D self-dual theory \cite{Fur} and in its promotion to four-dimensional space-time \cite{Bota}, \cite{Wo}, with the role of the Chern-Simons mass term played by the LV Carroll-Field-Jackiw (CFJ) term. The dual embedding of a four-dimensional Proca-like theory with a CPT-even Lorentz-breaking mass term was carried out in \cite{Scarp}, with the resulting theory involving very  interesting higher-derivative terms.
	
	At the same time, one aspect of duality discussed in \cite{scarp-epl} was lacking a deeper analysis. In that paper, a relation between the propagators of the two  models interrelated by duality was derived, and it is potentially dangerous for the theory constructed through the gauge embedding. Indeed, since the dual propagator of the gauge field looks like the difference between two propagators, one still faces the question whether the new model is contaminated by ghost modes of propagation of the vector field or not. Alternatives to avoid the emergency of ghosts in the process of dualization were developed \cite{Dalm-1}, \cite{Dalm-2}, \cite{Dalm-3}, \cite{Dalm-4}. In some approaches, the price to be paid is the lost of locality \cite{Dalm-1}. In this paper, we show that the complete analysis needs to take in consideration the new interaction terms which show up in the constructed model. It becomes evident that the traditional study of the saturated propagator should be improved to consider also the current-current interaction. When this additional contribution is considered, there occurs a cancellation of the spurious nonphysical terms. This conclusion is shown to be extended for quantum calculations.
	
	The structure of the paper looks like follows. In the section II, we perform the gauge embedding of our gauge-breaking model for a vector field, obtaining a new gauge invariant theory. In the section III, we compare propagators of these theories. In the section IV, we prove tree-level equivalence of the models, and, in the section V, we prove their perturbative equivalence at the one-loop order and discuss the situation at higher loops. Our results are discussed in section VI.

	\section{The gauge embedding procedure}
	\label{NDM}
	Let us consider the following Lagrangian density, which, in general, can represent itself as a Lorentz-violating extension of quantum electrodynamics (QED),
	\begin{equation}
	{\cal L}= {\cal L}_A + \frac 12 \mu^2 A^\mu M_{\mu\nu} A^\nu -J_\mu A^\mu + {\cal L}_F,
	\label{model}
	\end{equation}
	where ${\cal L}_A$ is the gauge invariant part of the photon sector, which can encompass CPT-odd or even Lorentz-breaking contributions, ${\cal L}_F$ is the free fermion Lagrangian density, $M_{\mu\nu}$ is a dimensionless tensor which depends on constant background tensors and $\mu$ is a parameter with the mass dimension 1.
	It is possible to construct, on the base of (\ref{model}), a gauge-invariant model  which describes the same physical solutions by means of a dualization method. Here, we use a gauge embedding technique also known as Noether dualization method (NDM), initially formulated in \cite{AnaWo} and further applied in some other papers, f.e. \cite{Fur}. 
	
	The first step is the calculation of the first variation of the Lagrangian density,
	\begin{equation}
	\delta {\cal L}[A_{\mu}]= K_\nu \, \delta A^\nu,
	\label{var-1}
	\end{equation}
	with $K^\mu$ the Noether current, so that we construct the first iterated Lagrangian by introducing an auxiliary field $B^\mu$,
	\be
	{\cal L}^{(1)} = {\cal L}-K^\mu B_\mu.
	\ee
	For $B^\mu$ transforming as $\delta B_{\mu}=\delta A_{\mu} = \partial_{\mu} \eta$, we have the following variation of the first-iterated Lagrangian:
	\begin{equation}
	\delta\,{\cal L}^{(1)}\,=\,-(\delta\,K_{\mu})\,B^{\mu}.
	\end{equation}
	Since the breaking of gauge symmetry occurs exclusively due to the mass term, i.e. $\delta{\cal L}_A=0$, we have
	\begin{equation}
	\delta K^\mu=\mu^2 M^{\mu \nu} \delta A_\nu,
	\end{equation}
	and
	\begin{equation}
	\label{varl1}
	\delta {\cal L}^{(1)}=-\mu^2 B_\mu M^{\mu \nu} \delta A_\nu.
	\end{equation}
	If we define the second iterated Lagrangian as
	\begin{equation}
	{\cal L}^{(2)}\,=\,{\cal L}^{(1)}\,+\,\frac{\mu^2}{2}\,B^{\mu}M_{\mu \nu}B^\nu
	\end{equation}
	and use the variation of $B_\mu$ and (\ref{varl1}), we find that its total variation vanishes, $\delta {\cal L}^{(2)}\,=\,0$. Let us write down the explicit form of this action,
	\begin{equation}
	{\cal L}^{(2)}= {\cal L} - K_\mu B^\mu + \frac{\mu^2}{2}\,B^{\mu}M_{\mu \nu}B^\nu.
	\label{master}
	\end{equation}
	After carrying out the variation of this action with relation to $B_\mu$, we get the following equation of motion:
	\begin{equation}
	K_\mu-\mu^2M_{\mu \nu} B^\nu=0.
	\end{equation}
	Plugging this back into (\ref{master}), we obtain the dual gauge invariant theory:
	\begin{equation}
	{\cal L}_D= {\cal L}-\frac{1}{2\mu^2}K^\mu L_{\mu\nu} K^\nu,
	\label{Dual-1}
	\end{equation}
	in which $L_{\mu\nu}=(M^{-1})_{\mu\nu}$. This is a general relation between the Lagrangians of the gauge-breaking model and its dual gauge-invariant theory, for the case when the mass term is the unique one responsible for the violation of this symmetry.
	
	\section{Relation between propagators and new interaction terms}
	
	Now let us obtain simple relations between the wave operators of the models involved in the duality relation (see, for example, \cite{scarp-epl}). The Lagrangian density for the gauge-breaking model can be written as
	\begin{equation}
	{\cal L}=\frac 12 A^\mu {\cal O}_{\mu\nu}A^\nu -J^\mu A_\mu + {\cal L}_F,
	\end{equation}
	being ${\cal O}_{\mu\nu}$ its wave operator. The variation of the above expression is given by
	\begin{equation}
	\delta {\cal L}=\left({\cal O} A - J\right) \delta A,
	\label{var-2}
	\end{equation}
	where, for compactness, we suppressed the Lorentz indices. The equation above was obtained after partial integrations in the action. For this, it is important to note that this differential operator ${\cal O}$ involves only terms of first and second orders in derivatives. We require the first-order term to be antisymmetric in the Lorentz indices, as it occurs in the Chern-Simons case.  
	
	Comparing equations (\ref{var-1}) and (\ref{var-2}), we get
	\be
	K={\cal O}A - J,
	\ee
	such that we have
	\be
	\frac{1}{2\mu^2}K L K=\frac {1}{2 \mu^2}\left({\cal O}A - J\right) L \left({\cal O}A - J\right).
	\ee
	The same argument used before is valid now for new integrations by parts in the action, which will provide us with the result
	\be
	\frac{1}{2\mu^2}K L K=\frac 12 A \left(\frac {1}{\mu^2}{\cal O}L\cal{O} \right)A
	- \frac{1}{\mu^2}J\,L({\cal O}A) + \frac{1}{2\mu^2}J\,L\,J.
	\ee
	Now, let us consider a model with the quadratic part $\frac 12 A{\cal O}_0 A$, such that ${\cal O}_0={\cal O}-\mu^2 M$, which is simply the wave operator of our model without the mass term. With this definition, we have
	\be
	{\cal O}L{\cal O}={\cal O}L\left({\cal O}_0 +\mu^2 M\right)={\cal O}L{\cal O}_0 + \mu^2 {\cal O}.
	\ee
	When we plug this result back into equation (\ref{Dual-1}), we get
	\be
	{\cal L}_D=\frac 12 A\left(-\frac{1}{\mu^2}{\cal O}L {\cal O}_0 \right)A + {\cal L}'_I+ {\cal L}'_F,
	\label{waveD}
	\ee
	with
	\be
	{\cal L}'_I=-J\,A + \frac{1}{\mu^2}J\,L\,({\cal O}A)
	\ee
	and
	\be
	{\cal L}'_F={\cal L}_F - \frac{1}{2\mu^2}J\,L\,J.
	\ee
	From (\ref{waveD}), we identify the wave operator for the dualized model as
	\be
	{\cal O}_D=-\frac{1}{\mu^2}{\cal O}L {\cal O}_0,
	\label{Dwave}
	\ee
	which is an interesting relation between the wave operators of the two models. As we will see, this has strong implications when the field dynamics is studied. It is also worth to emphasize that the higher order derivative terms will bring new modes of field propagation which may affect the physical equivalence between the dual models. Besides, it is important to note that new interaction terms were generated: a nonminimal one, responsible for a magnetic-like interaction; and a current-current Thirring-like coupling.
	
	Let us now take care of the quadratic part of the dualized gauge action. Equation (\ref{Dwave}) allows a simple procedure for the calculation of the gauge propagator of the dualized model. First, a gauge-fixing term is necessary, since ${\cal O}_0$ has no inverse. But here we have two possibilities: the gauge-fixing can be added to ${\cal O}_0$ or to ${\cal O}_D$. In the first situation, we use
	\be
	\tilde{{\cal O}}_0=  {\cal O}_0 +\frac{\Box}{\tau}\omega
	\ee
	in the place of ${\cal O}_0$, where $\tau$ is the gauge-fixing parameter and $\omega_{\mu\nu}=\partial_\mu \partial_\nu/\Box$ is the longitudinal spin projector. The gauge-fixed wave operator of the dualized model is then given by
	\bq
	{\cal O}_D&=& -\frac{1}{\mu^2}{\cal O}L \tilde{{\cal O}}_0 \nonumber \\
	&=& -\frac{1}{\mu^2}{\cal O}L {\cal O}_0 + \frac{\Box}{\tau} \left(-\frac{1}{\mu^2}{\cal O}\,L\, \omega \right).
	\eq
	Effectively, we have a new gauge-fixing term in which the projector $\omega_{\mu\nu}$ is replaced by another longitudinal operator $-\frac{1}{\mu^2}\left({\cal O}\,L\, \omega \right)_{\mu\nu}$. This way of dealing with the problem is very interesting, as it allows a very simple calculation of the propagator for the vector field, which is given by
	\be
	\left<A_\mu A_\nu\right>_D=i \left({\cal O}_D^{-1}\right)_{\mu\nu}.
	\ee
	The inversion of the dual wave operator is straightforward, so that we have
	\be
	{\cal O}_D^{-1}=-\mu^2 \tilde{\cal O}_0^{-1}\, M \, {\cal O}^{-1}.
	\ee
	It is possible to obtain a simpler relation. First, we have
	\be
	{\cal O}=\tilde{\cal O}_0 + \mu^2 M - \frac{\Box}{\tau}\omega,
	\ee
	from which, after making the product of $\tilde{\cal O}_0^{-1}$, on the right, and of ${\cal O}^{-1}$, on the left, we obtain
	\be
	\tilde{\cal O}_0^{-1}={\cal O}^{-1} + \mu^2 \tilde{\cal O}_0^{-1}\, M \, {\cal O}^{-1} 
	- \frac{\Box}{\tau} \tilde{\cal O}_0^{-1} \, \omega \, {\cal O}^{-1}.
	\label{prop-relation-1}
	\ee
	We then identify the second term on the right hand side as $-{\cal O}_D^{-1}$ and, in the third, we write
	\be
	\tilde{\cal O}_0^{-1} = \tilde{\cal O}_{0\,Tr}^{-1}+ \frac{\tau}{\Box}\omega,
	\label{transversal}
	\ee
	where $\tilde{\cal O}_{0\,Tr}^{-1}$ is the transversal part of $\tilde{\cal O}_0^{-1}$. This decomposition is explained by the fact that the model which originates the wave operator ${\cal O}_0$ is gauge invariant. So, we have
	\bq
	- \frac{\Box}{\tau} \tilde{\cal O}_0^{-1} \, \omega \, {\cal O}^{-1}&=& - \frac{\Box}{\tau}
	\left(\tilde{\cal O}_{0\,Tr}^{-1}+ \frac{\tau}{\Box}\omega \right) \, \omega \, {\cal O}^{-1} \nonumber \\
	&=& - \omega {\cal O}^{-1} = \theta {\cal O}^{-1} - {\cal O}^{-1},
	\eq
	in which $\theta_{\mu\nu}= \eta_{\mu\nu}-\omega_{\mu\nu}$ is the transversal spin operator projector. If we use the above relation in equation (\ref{prop-relation-1}), we get
	\be
	{\cal O}_D^{-1}= \theta {\cal O}^{-1} - \tilde{\cal O}_0^{-1}.
	\label{prop-relation-2}
	\ee
	
	The relation between propagators shown in Eq. (\ref{prop-relation-2}) is remarkably simple and more general than the one obtained in \cite{scarp-epl}. It is worth to note that it does not depend on the explicit form of $M_{\mu\nu}$. In addition, the expression makes clear the presence of the same modes of propagation as the original theory, which are naturally assumed to describe the same particles. Earlier, the arising of modes characterized by the same dispersion relations has been observed in \cite{Fur}, \cite{sdual-even}, for three-dimensional LV self-dual and Maxwell-Chern-Simons (MCS) models. However, attention should be paid to the presence of new poles in the propagator, coming from the massless theory. As they appear with the sign exchanged in the expression, this may indicate the presence of ghost modes in the dual model. This possibility should be verified in detail. 
	
	\section{Tree level equivalence}
	
	In the previous section, we derived a relation between the propagators of the original model and its dual gauge invariant partner. From this relation, which is a difference of two propagators, there emerges the dangerous possibility of nonphysical modes, ghosts,  which could spoil the energy spectrum of the model. In this section, we show that the traditional analysis of the saturated propagator, similar to that one performed, e.g. in \cite{Scarp}, needs, in this case, to be extended by the inclusion of the current-current interaction, which will cause the exact cancellation of the non-physical contributions. 
	
	Let us write the dual Lagrangian density in the following form
	\be
	{\cal L}_D= \frac 12 A^\mu {\cal O}_{D\mu\nu}A^\nu -\left[J_\mu - \frac{1}{\mu^2}J_\alpha L^{\alpha\beta}{\cal O}_{\beta \mu}\right]A^\mu 
	- \frac{1}{2\mu^2} J_\mu L^{\mu\nu}J_\nu + {\cal L}_F,
	\ee
	from which we define the new current as
	\be
	\tilde{J}_\mu = J_\nu \left(\delta_\mu^\nu -  \frac{1}{\mu^2} L^{\nu\beta}{\cal O}_{\beta \mu}\right),
	\ee
	which is easily shown to be conserved. In the momentum space, we have
	\be
	p^\mu \tilde{J}_\mu=p^\mu\left(J_\mu - \frac{1}{\mu^2}{\cal O}_{\mu\beta} L^{\beta\nu}J_\nu \right),
	\ee
	and, since ${\cal O}$ includes a transverse part and the $M$ dependent one, we write
	\be
	- \frac{1}{\mu^2}p^\mu{\cal O}_{\mu\beta} L^{\beta\nu}J_\nu= - \frac{1}{\mu^2}p^\mu({\cal O}_{0\mu\beta }+\mu^2 M_{\mu\beta}) L^{\beta\nu}J_\nu = -p^\nu J_\nu,
	\ee
	where we have used the fact that $L=M^{-1}$ and that ${\cal O}_0$ is transverse (the original massless model is gauge-invariant). We then have $p^\mu \tilde{J}_\mu=0$, a relation which will be useful in our demonstration. 
	
	We will apply the electron-electron scattering in order to show the tree-level equivalence between the two models. For the original gauge-violating model, the amplitude is represented by the graph depicted in Fig. 1 added to its crossed diagram. The on-shell amplitude corresponding to Fig. 1 is given by 
	\be
	F=-i J {\cal O}^{-1} J,
	\ee
	in which, again, we omitted the Lorentz indices.
	\begin{center}
		\begin{picture}(300,100)(0,20)
		\Vertex(150,70){2} \Vertex(150,50){2}
		\Photon(150,70)(150,50){3}{3}
		\ArrowLine(150,70)(130,90) \ArrowLine(170,90)(150,70)
		\ArrowLine(150,50)(130,30) \ArrowLine(170,30)(150,50)
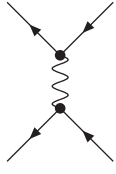
		\end{picture} \\ {\sl FIG. 1. Electron-electron scattering for the original massive model.}
	\end{center}
	The same calculation, when performed for the dual model, must also include the contribution of the tree graph coming from the four-fermion vertex. This contribution is presented graphically in Fig. 2, which does not show the crossed diagrams.
	\begin{center}
		\begin{picture}(300,100)(0,0)
		\Vertex(80,70){2} \Vertex(80,50){2}
		\Photon(80,70)(80,50){3}{3}
		\ArrowLine(80,70)(60,90)\ArrowLine(100,90)(80,70)
		\ArrowLine(80,50)(60,30) \ArrowLine(100,30)(80,50) \Text(80,20)[t]{$(a)$}
		\ArrowLine(220,60)(200,90) \ArrowLine(240,90)(220,60) \Vertex(220,60){2}
		\ArrowLine(220,60)(200,30) \ArrowLine(240,30)(220,60) \Text(220,20)[t]{$(b)$}
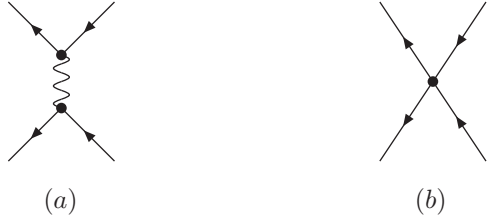
		\end{picture}\\ {\sl FIG. 2. Contributions to the fermion-fermion scattering for the dual model.}
	\end{center}
	We then have, for the dual model,
	\be
	\tilde{F}= - i \tilde{J} {\cal O}_D^{-1} \tilde{J} - i J \frac{L}{\mu^2} J.
	\ee
	Let us calculate the first term. Since the current $\tilde{J}$ is conserved, we can discard terms of ${\cal O}_D^{-1}$ which produces contractions of the photon momentum with the modified current yielding zero result. We can then use the identity
	\bq
	{\cal O}_D^{-1}&=& \theta{\cal O}^{-1}- \tilde{\cal O}_o^{-1} \nonumber \\
	&=&  {\cal O}^{-1}- \tilde{\cal O}_o^{-1} - \omega {\cal O}^{-1} \to {\cal O}^{-1}- \tilde{\cal O}_o^{-1}
	\eq
	in our calculation, in which we took into account that $\theta + \omega = \eta$. So, we write
	\bq
	- i \tilde{J} {\cal O}_D^{-1} \tilde{J} &=& -iJ\left(\eta -\frac{1}{\mu^2}L {\cal O}\right) 
	\left({\cal O}^{-1}- \tilde{\cal O}_o^{-1}\right)\left(\eta -\frac{1}{\mu^2}{\cal O}L\right)J \nonumber \\
	&=& -i J {\cal O}^{-1} J + \frac{i}{\mu^2}JLJ + iJ \tilde{\cal O}_0^{-1}J 
	-\frac{i}{\mu^2}J\tilde{\cal O}_0^{-1}{\cal O}LJ + \nonumber \\
	&+& \frac{i}{\mu^2} JLJ -\frac{i}{\mu^4}JL{\cal O}LJ -\frac{i}{\mu^2}JL{\cal O}\tilde{\cal O}_0^{-1}J 
	+ \frac{i}{\mu^4}JL{\cal O}\tilde{\cal O}_0^{-1}{\cal O}LJ.
	\label{expansion}
	\eq
	We now substitute the relation ${\cal O}=\tilde{\cal O}_0 + \mu^2 M + \frac{p^2}{\tau} \omega$ into the last two terms of Eq. (\ref{expansion}) to obtain
	\bq
	-\frac{i}{\mu^2}JL{\cal O}\tilde{\cal O}_0^{-1}J&=& - \frac{i}{\mu^2}J \left[L + \tilde{\cal O}_0^{-1} + \frac {p^2}{\tau}L \omega \tilde{\cal O}_0^{-1} \right]J; 
	\label{38}\\
	\frac{i}{\mu^4}JL{\cal O}\tilde{\cal O}_0^{-1}{\cal O}LJ &=& \frac{i}{\mu^4}J\left[L{\cal O}L + \mu^2 \tilde{\cal O}_0^{-1}{\cal O}L + \frac{p^2}{\tau} L \omega L +\frac{p^2}{\tau}\mu^2 L \omega \tilde{\cal O}_0^{-1} + \frac{p^4}{\tau^2} L \omega \tilde{\cal O}_0^{-1} \omega L \right]J. \label{39}
	\eq
	We then use (\ref{transversal}) in the momentum space, i.e. $\tilde{\cal O}_0^{-1} = \tilde{\cal O}_{0\,Tr}^{-1}- \frac{\tau}{p^2}\omega$, in the last term of (\ref{39}), and the relations $\omega \tilde{\cal O}_{0\,Tr}^{-1}=0$ and $\omega^2=\omega$ to write
	\be
	\frac{p^4}{\tau^2} L \omega \tilde{\cal O}_0^{-1} \omega L = -\frac{p^2}{\tau}L \omega L,
	\ee
	so that 
	\be
	\frac{i}{\mu^4}JL{\cal O}\tilde{\cal O}_0^{-1}{\cal O}LJ =\frac{i}{\mu^4}J\left[L{\cal O}L + \mu^2 \tilde{\cal O}_0^{-1}{\cal O}L +  \frac{p^2}{\tau}\mu^2 L \omega \tilde{\cal O}_0^{-1} \right]J
	\label{41}.
	\ee
	After substituting (\ref{38}) and (\ref{41}) in (\ref{expansion}), we finally get
	\be
	- i \tilde{J} {\cal O}_D^{-1} \tilde{J} = -i J {\cal O}^{-1} J + \frac{i}{\mu^2}JLJ,
	\label{result}
	\ee
	such that $F=\tilde{F}$. This is a on-shell result, which is maintained for the crossed diagrams. In the next section we will show with an one-loop example that this equivalence remains. Besides, we will show this is also true for the off-shell calculation.

	\section{Equivalence at higher loop orders}
	The equivalence we showed in the previous section is an on-shell result. To generalize it, in this section, we first demonstrate the on-shell equivalence for a simple one-loop amplitude and, then, we show the equivalence is valid off-shell and argue that this implies that the result can be extended to all loop orders. Let us begin by considering the electron self-energy of some extended massive electrodynamics and of its corresponding gauge-invariant dual model. The amplitude for the massive model, which is graphically represented in Fig. 3, is given by
	\be
	i\Sigma(p)=-q^2 \int^\Lambda\frac{d^4 k}{(2\pi)^4} \frac{\gamma^\mu (\pslash -\kslash + m)\gamma^\nu}{[(p-k)^2-m^2]} ({\cal O}^{-1})_{\mu\nu}(k).
	\ee
	
	On the other hand, the fermion self-energy for the dual model receives the contributions from the graphs (a) and (b) of Fig. 4,
	\be
	i \tilde{\Sigma}(p) = i \tilde{\Sigma}_a(p) + i \tilde{\Sigma}_b(p),
	\ee
	for which we have
	\be
	i\tilde{\Sigma}_a(p) = ´-q^2 \int^\Lambda\frac{d^4 k}{(2\pi)^4} \frac{\gamma^\mu \left(\eta -\frac{1}{\mu^2}L {\cal O}\right)_{\mu\alpha} (\pslash -\kslash + m)\left(\eta -\frac{1}{\mu^2} {\cal O}L\right)_{\beta\nu}\gamma^\nu}{[(p-k)^2-m^2]} ({\cal O}^{-1})_D^{\alpha\beta}(k)
	\ee
	and
	\be
	i\tilde{\Sigma}_b(p) = ´\frac{q^2}{\mu^2} \int^\Lambda\frac{d^4 k}{(2\pi)^4} \frac{\gamma^\mu L_{\mu\nu}\gamma^\nu(\kslash + m)}{[k^2-m^2]}=´\frac{mq^2}{\mu^2} \int^\Lambda\frac{d^4 k}{(2\pi)^4} \frac{\gamma^\mu L_{\mu\nu}\gamma^\nu}{[k^2-m^2]}.
	\label{product}
	\ee
	\begin{center}
		\begin{picture}(300,100)(0,20)
		\Vertex(180,50){2} \Vertex(120,50){2}
		\Line(90,50)(210,50) 
		\LongArrow(140,55)(160,55) 	\LongArrow(100,55)(110,55) \LongArrow(190,55)(200,55)
		\LongArrowArcn(150,50)(20,120,60) 
		\PhotonArc(150,50)(30,0,180){4}{8.5}  \Text(150,40)[t]{$i\Sigma(p)$} 
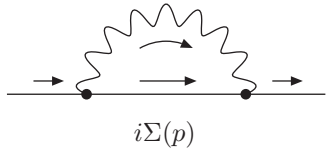
		\end{picture} \\ {\sl FIG. 3. Self-energy diagram for the fermion in the original massive model.}
	\end{center}
	
	\begin{center}
		\begin{picture}(300,100)(0,20)
		\Vertex(110,50){2} \Vertex(50,50){2}
		\Line(20,50)(140,50) 
		\LongArrow(70,55)(90,55) 	\LongArrow(30,55)(40,55) 	\LongArrow(120,55)(130,55)
		\LongArrowArcn(80,50)(20,120,60) 
		\PhotonArc(80,50)(30,0,180){4}{8.5}  \Text(80,40)[t]{$(a)\,\,\,\,i\tilde{\Sigma}_a(p)$} 
		\CArc (220,70)(20,0,360) \Text(220,40)[t]{$(b)\,\,\,\,i\tilde{\Sigma}_b(p)$}  
		\LongArrowArcn(220,70)(15,130,50)
		\Line(160,50)(280,50) \Vertex(220,50){2} \LongArrow(185,55)(195,55) 	\LongArrow(245,55)(255,55)
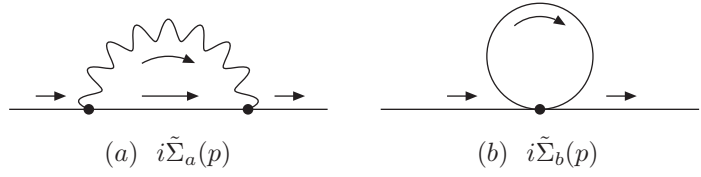
		\end{picture} \\ {\sl FIG. 4. Contributions to the one-loop self-energy of the fermion in the dual model.}
	\end{center}
	
	Note that the operator ${\cal O}$ acts on the vector field and, therefore, depends only on the momentum $k^\mu$. Let us consider first the amplitude $i\tilde{\Sigma}_a(p)$. We have to deal with the product of operators
	\be
	\left(\eta -\frac{1}{\mu^2}L {\cal O}\right)_{\mu\alpha}({\cal O}^{-1})_D^{\alpha\beta}(k)\left(\eta -\frac{1}{\mu^2} {\cal O}L\right)_{\beta\nu}.
	\ee
	In the tree level calculation of the last section, we disregarded the term in $\omega^{\alpha\beta}$ in $({\cal O}^{-1})_D^{\alpha\beta}$ due to the conservation of the current $\tilde{J}$, $k^\mu \tilde{J}_\mu=0$. Let us consider this term here and show how the on-shell calculation fixes this contribution to be zero. The possible non-transverse parts of the vertices are proportional to $\omega$. Then, since $\omega^2=\omega$ because $\omega$ is a projector, the contribution due to the disregarded term, in the numerator, involves the factor
	\be
	\gamma^\mu \omega_{\mu\nu} (\pslash -\kslash + m)\gamma^\nu = \frac{1}{k^2}\kslash (\pslash -\kslash + m) \kslash,
	\ee
	and is proportional to
	\be
	I=\int^\Lambda\frac{d^4 k}{(2\pi)^4} \frac{\kslash (\pslash -\kslash + m) \kslash}{k^2[(p-k)^2-m^2]}.
	\ee
	In order to calculate $I$, we use Implicit Regularization (IReg) methodology described in \cite{IR-Bruno}. Assuming the integral is regularized, we perform Feynman parametrization to obtain
	\be
	I= \int_0^1 \, dx\int^\Lambda\frac{d^4 k}{(2\pi)^4} \frac{(\kslash + x \pslash)[(1-x)\pslash -\kslash + m] (\kslash + x \pslash)}{(k^2+H^2)^2},
	\ee
	with $H^2=p^2x(1-x)-m^2x$. Selecting the even terms in the numerator and using the on-shell conditions $\bar{u}(p)\pslash =\bar{u}(p) m$, $\pslash u(p) = m u(p)$ and $p^2=m^2$, we get
	\bq
	I&=&\int_0^1 \, dx \left\{-mx \int^\Lambda\frac{d^4 k}{(2\pi)^4}\frac{1}{(k^2+H^2)} + 2(1-x)p^\alpha \gamma^\beta \int^\Lambda\frac{d^4 k}{(2\pi)^4}\frac{k_\alpha k_\beta}{(k^2+H^2)^2} + \right. \nonumber \\
	&+& \left. \left[m x H^2 + m^3 x^2 (2-x) \right] \int^\Lambda\frac{d^4 k}{(2\pi)^4}\frac{1}{(k^2+H^2)^2}\right\}.
	\eq
	Now we use the relation
	\be
	\int^\Lambda\frac{d^4 k}{(2\pi)^4}\frac{k_\alpha k_\beta}{(k^2+H^2)^2} = \frac{\eta_{\alpha\beta}}{2} I_{quad}(-H^2)
	\ee
	with the aim of eliminating the surface terms, which is a condition for gauge invariance, and the on-shell result $H^2=-m^2x^2$, to obtain
	\be
	I=\int_0^1 \, dx \left\{(1-2x)m I_{quad}(-H^2) + 2m^3x^2(1-x)I_{log}(-H^2) \right\},
	\ee
	where we use the definitions of the basic divergences of IReg,
	\be
	I_{log}(m^2) = \int^\Lambda\frac{d^4 k}{(2\pi)^4}\frac{1}{(k^2-m^2)^2}
	\ee
	and
	\be
	I_{quad}(m^2)= \int^\Lambda\frac{d^4 k}{(2\pi)^4}\frac{1}{(k^2-m^2)}.
	\ee
	Next, the following scale relations are used in order to have the basic divergences free from the external momentum,
	\be
	I_{log}(-H^2)= I_{log}(m^2) - \frac{i}{16\pi^2}\ln{\left(-\frac{H^2}{m^2}\right)}
	\label{scale-1}
	\ee
	and
	\be
	I_{quad}(-H^2)=I_{quad}(m^2)-(m^2+H^2)I_{log}(m^2) - \frac{i}{16\pi^2} 
	\left[m^2 + H^2 - H^2 \ln{\left(-\frac{H^2}{m^2}\right)} \right].
	\label{scale-2}
	\ee
	We get
	\bq
	I &=& I_{quad}(m^2)\int_0^1 \,dx\, m (1-2x) + I_{log}(m^2)\int_0^1\, dx \, m^3(1-x)(4x^2+x-1) + \nonumber \\
	&-& \frac{i}{16 \pi^2} \int_0^1 \, dx \, m^3 \left[ (1-2x)(1-x^2) + x^2(3-4x) \ln{x^2}\right] = 0.
	\eq
	
	This shows that even in the presence of a non-transverse vertex in the dual model this spurious term does not contribute to physical calculations. However, as it was already indirectly shown in the previous section, we have
	\be
	\left(\eta -\frac{1}{\mu^2}L {\cal O}\right)^{\mu\nu}k_\nu
	= k^\mu - \frac{1}{\mu^2}\left[L({\cal O}_0 + \mu^2 M)\right]^{\mu\nu}k_\nu=0,
	\ee
	since ${\cal O}_0$ is transverse and $M=L^{-1}$. We then turn our attention to the Eq. (\ref{product}). Given that the vertices are transverse, we follow exactly the same steps of the calculations from Eq. (\ref{expansion}) to Eq. (\ref{result}), to get
	\be
	\left(\eta -\frac{1}{\mu^2}L {\cal O}\right)_{\mu\alpha}({\cal O}^{-1})_D^{\alpha\beta}(k)\left(\eta -\frac{1}{\mu^2} {\cal O}L\right)_{\beta\nu}
	= \left({\cal O}^{-1} + \frac{1}{\mu^2}L \right)_{\mu\nu}.
	\label{of-shell-tree}
	\ee
	As a consequence, we can write
	\be
	i \tilde{\Sigma}_a= -q^2 \int^\Lambda\frac{d^4 k}{(2\pi)^4} \frac{\gamma^\mu (\pslash -\kslash + m)\gamma^\nu}{[(p-k)^2-m^2]} ({\cal O}^{-1})_{\mu\nu} 
	- \frac{q^2}{\mu^2} \int^\Lambda\frac{d^4 k}{(2\pi)^4} \frac{\gamma^\mu (\pslash -\kslash + m)\gamma^\nu L_{\mu\nu}}{[(p-k)^2-m^2]}.
	\ee
	The first term is $i\Sigma(p)$ and, in the second, we make the shift $k \to k + p$ and discard the surface term, as prescribed by IReg. The amplitude reads
	\bq
	i \tilde{\Sigma}_a &=& i \Sigma - \frac{q^2}{\mu^2} \int^\Lambda\frac{d^4 k}{(2\pi)^4} \frac{\gamma^\mu (-\kslash + m)\gamma^\nu L_{\mu\nu}}{(k^2-m^2)} \nonumber \\
	&=& i \Sigma - \frac{mq^2}{\mu^2} \int^\Lambda\frac{d^4 k}{(2\pi)^4} \frac{\gamma^\mu \gamma^\nu L_{\mu\nu}}{(k^2-m^2)}
	= i \Sigma - i \tilde{\Sigma}_b,
	\eq
	and we have
	\be
	i \tilde{\Sigma} = i \tilde{\Sigma}_a + i \tilde{\Sigma}_b = i \Sigma.
	\ee
	
	We emphasize that this result takes place off-shell. It is natural to expect that the result of equation (\ref{of-shell-tree}) assures that the equivalence between the two models interrelated through the dualization process can be extended to all loop orders. This occurs since every tree diagram of the type shown in Figure 1, referring to the original massive model, which composes an amplitude, will correspond a pair of diagrams of the types shown in Figure 2, referring to the dual gauge invariant model. As a result of this, we hope that it will be possible to construct a diagrammatic proof of the equivalence to all loop orders between the models interconnected by dualization. For example, for a pair of graphs of $(n-1)$ loop order that are correlate, and that have $l$ external legs of fermions, it is possible, by joining two of them, to construct a new pair of diagrams of $n$ loop order and $l-2$ external fermion legs. We expect that an all-loop order equivalence proof can be carried out using a procedure similar to the subtraction of divergences and subdivergences with the use of the forests formula mathematically established by the Bogoliubov– Parasiuk–Hepp–Zimmermann (BPHZ) theorem \cite{BPHZ-1}, \cite{BPHZ-2}. The quartic vertices of the gauge-invariant dual model will play the role of collapsed subgraphs in ``counterterms'' to cancel out the spurious terms.

	\section{Summary}
	
	We studied the gauge embedding methodology for a completely generic gauge-breaking model of the vector field, which can be defined in a spacetime of an arbitrary dimension, can include higher derivatives or not, is Lorentz-invariant or Lorentz-breaking, etc. This methodology, as usual, implies in the arising of a new gauge invariant model for the vector field. We obtained the relationship between propagators of these models, which, in particular, implies that they display common modes, describing therefore the same particles (earlier a similar study has been performed for a certain LV extension of the three-dimensional self-dual model and the dual theory representing itself as some extension of the MCS theory in \cite{Fur}). Then, we demonstrated that the duality also holds for tree-level scattering amplitudes and one-loop contributions, and, moreover, can be naturally expected to occur for loop corrections of any order. Effectively, in this paper we not only presented a prescription for obtaining the dual gauge invariant theory for an arbitrary gauge-breaking one, but also argued that the duality keeps both at classical and quantum levels.
	
	Now, let us briefly discuss possible extensions of our studies. First of all, we hope that it will be possible to construct a diagrammatic proof of the equivalence to all-loop orders between the models interconnected by dualization, by using a procedure similar to that prescribed by the forests formula of the BPHZ theorem. It is also interesting to investigate the relation between this kind of duality and the equivalence theorems described in \cite{Salam}. Besides, we intend, in forthcoming papers, to perform a better study of duality in noncommutative field theory models, generalizing results of \cite{NC} and investigate the extension of the concept of duality to higher spin theories, especially, to gravitational ones.

	\vspace*{1mm}
	
	{\bf Acknowledgments.} Authors are grateful to J. R. Nascimento for important discussions. A. P. B. S. acknowledges CNPq for financial support. The work by A. Yu. P. has been partially supported by the CNPq project No. 301562/2019-9.

\end{document}